# Temperature dependence of the ultra-cold neutron's loss rate

S.Arzumanov[1], J.Butterworth[2], R.Golub[3], P.Geltenbort[2], E.Korobkina[3,aa]

[1] RRC Kurchatov Institut, 1 Kurchatov sq., 123182 Moscow, Russia
[2] Institute Laue-Langevin, BP156, 38042 Grenoble Cedex 9, France
[3] Hahn-Meitner-Institute, Glienikerstr.100, 14109 Berlin, Germany

**Abstract**

We report about the first results obtained with our Ultra High Vacuum cryostat recently commissioned at HMI to study Ultra Cold Neutrons(UCN) loss rate under low temperatures(down to 4K) and well controlled sample environment. Our analysis of the temperature dependence of UCN loss rate on Cu surface allowed us to rule out the subbarier model in support of the 1/v model and propose a reasonable explanation of the low temperature upscattering.

PACs: 61.12.-q Neutron diffraction and scattering 61.12.Ex Neutron scattering (including small-angle scattering)- 61.12.Bt Theories of diffraction and scattering-

**Introduction**

The inelastic upscattering on the hydrogen is known to be the main reason for UCN losses at room temperature. The main evidences are

- direct observation of Hydrogen in the surface 50A layers of Cu and Carbon in the range of (50-30)% after applying the usual surface treatment technique used in UCN storage bottles[1].
- study of the UCN loss on samples by detecting upscattered neutrons[2]
- Observation of the temperature dependence of the UCN loss rate [2,3,4].
- study of UCN capture by Hydrogen and upscattering using UCN prompt γ-analysis shows a linear dependence of the probability of inelastic scattering on the amount of Hydrogen[5,14]; moreover, this technique allowed to measure the absolute value of the upscattering cross section, which was found to be in the range 5-7 barn per proton for different materials.

Nevertheless, we can not say that UCN upscattering is well studied and understood. Indeed, the experimental data are rather poor, especially in the most interesting low temperature range, i.e. below 77K.

Practically, there are only two studies of the temperature dependence down to 6.5K [3] and 13.5K [4]. The results of both at first glance seem to give evidence of excess losses at low temperature and give rise to the idea of a low temperature "anomaly" in UCN interaction with solid matter. The common feature of both studies is that UCN were stored in bottles coated with Be. Therefore it seems logical to address the issue of the quality of the surface. Indeed, the dependence of the loss rate on the energy of stored UCN in [4] looks more like losses on Al than on Be. Our test of Be and double Be(Stainless steel) coatings on Al foils and Si wafer[6] directly proved that it is really a case. The more rigorous study of Be coatings also gave evidence that the transmission can not be made less than $5 \cdot 10^{-5}$ [7]. The energy dependency of the transmission shows that it arise due to pinholes in the Be coating. Since the losses are energy dependent, presence of pinholes deformed spectrum different way for the low and higher energies making it impossible to derive a correct probability of the losses. Therefore, the data from [3,4] could not be used for reliable analysis.

---

[a] Corresponding author, e-mal: korbkina@hmi.de, tel.+49/(0)30 80622290

Recently, a more detail studies of the temperature dependence on several materials was performed but only down to 77K. The conclusion of study of the Be foil was that at 77K there is still loss probability equal to $2.5 \cdot 10^{-5}$ per collision[8]. The theoretical model (hydrogen oscillation with phonon frequency of the Be spectrum), giving a reasonable value for the cross section at 300K, was not able to reproduce the temperature dependence below 300K. Calculating a loss probability per collision, authors have got agreement only when taking 96% of Hydrogen concentration in the surface layer, which seems to be unreasonable. The results of [8] on Be were the opposite of an earlier measurement in Dubna [9], where the upscattering at 77K was found to be negligible. Study of pyrolytic graphite and fomblin oil coating again shows a considerable loss rate at 77K [10]. Authors assumed that the origin is related to bound conditions of Hydrogen. A small admixture of 2D-gas on the surface might easily explain such losses but Hydrogen gas is known to be physadsorbed only at temperatures below 30K.

Roughly speaking, it is clear that UCN are upscattered by Hydrogen, but it is not clear how and on what kind of Hydrogen – chemically tightly or weakly bound to the bulk material or on the 2D-gas, or something else. The recent result of the neutron life time measurement on the Fomblin oil again addressed the issue of the low temperature losses to the upscattering cross sections [11], which was found to be several times higher that expected[12].

Neither the theoretical study of the upscattering has been done at the proper basis. The model calculations of UCN upscattering exist only on the rather naïve estimations level. The first consistent approach taking into account the subbarrier nature of the process is under development[13]. The first results for the coherent scattering predict an additional effect from the very low energy relaxations but the relative value is rather small as well as all coherent effects. The next step will be incoherent effects on Hydrogen.

To test the theoretical models a study of the samples with known properties are required. For theoretical interpretation, it is more convenient to use pure monoelement materials well studied with UCN. In this publication we present results of the first test measurement of our ultra high vacuum cryostat, which allows us to study the UCN loss rate on Copper in a well controlled environment over a wide temperature range. The experiment was performed on the PF2 instrument, test beam position at the Institute Laue-Langevin.

## Description of the Experiment

Since UCN are interacting with a thin surface layer (approximately 100Å) the main issue of an experimental set-up is to prevent any changing of the surface state during cooling or heating. When cooling down to 20K the absorption of residual hydrogen may occur. When heating up one should take care about possible adsorption from hotter outgasing surfaces. In experiments [3] and [10] this problem was solved by separating vacuums and sealing the storage volume after outgassing under heating while in [4] the storage bottle was enclosed in an outer housing that was not sealed but continuously pumped through 77K trap. It is the outer housing that was heated and cooled. Therefore, in principle one can not exclude that excess (relative to theory) losses in [4] at 4K may partly originate from both the heating procedure, when impurities come from the hot housing and/or cooling, when owing to the long time of the measurement some residual gas gets to be adsorbed through the pumping line. Moreover, as was recently observed [14], heating above 200° C may produce some phase or destructive changes on the surface, and the effect of UCN small heating gets to be enhanced by several orders of magnitude. In our study we tried to take care and keep the environment of our surface under control.

Our experimental layout is shown at Fig.1. We don't use a separate vacuum but the whole apparatus was constructed to Ultra High Vacuum standards, i.e. only metal seals were used (CF flanges with Cu gaskets and wire sealing from annealed Al, In and Au ); no plastic parts, only metal and ceramic; dry pump system (turbomolecular pump and scroll forpump, Varian).

Both heating and cooling were performed by the direct contact of the storage bottle with a central part of the cryostat, which contained either a heating resistor or LN or LHe. Thus during heating the storage bottle was the hottest part of the apparatus whereas during cooling and at room temperature the coolest part was a second cryostat filled with LN and LHe. Both cryostats are of the same construction (Oxford instrument), but the UCN cryostat is attached to the UCN bottle and in the cryopump cryostat is connected to large area Cu baffles.

The UCN storage bottle was surrounded by an additional 4K thermal shield made from Cu, attached to the top of the bottle and several layers of Al foils. One Pt100 (at the bottom) and two carbon resistors (bottom and top) were used to control the temperature. The temperature of the bottom of the 77K shield was also controlled by a Pt100. A third carbon resistor was mounted on the sliding UCN shutter.

Both the UCN bottle and 77K shield were made from unpolished Oxygen Free High Conductivity Copper (OFHC). The UCN bottle was a horizontal cylinder 18 cm inner diameter and 19 cm long. UCN entered the bottle through a 5 cm diameter opening in one of the vertical end flanges. A sliding shutter was used to close the entrance hole. The opening in the 77K shield has a 10 μm Al window to prevent 300K irradiation from the inner neutron guide affect the 4K part. The 300K vacuum housing was attached to the PF2 test beam through an adapter flange with 100 μm vacuum tight Al window. The neutron detector was attached to the guide switcher. The distance between UCN bottle and UCN detector was about 1.5 m.

It took us 25 minutes to cool the UCN bottle from 77K down to 4K. We had to refill the LHe bath twice per day. The LN filling was arranged to be automatic. The best vacuum during measurement was at the lowest limit of our vacuum gauge, i.e. $\leq 10^{-9}$ mbar.

We performed two runs for two different states of the Cu surface. Prior to the first run the bottle was ultrasound cleaned in distilled water and before the second run underwent deuteration. Before deuteration the surface of the bottle was etched with 5% nitric acid and washed with acetone and ethanol. Deuteration was performed by heating in a vacuum oven up to 240°C and slow cooling down to room temperature in the vapour of $D_2O$ at approximately 10 mbar pressure during 24 hours.

The sequence of data taking for both surface states was as follows:
1   300K, room temperature, initial state;
2   450K, steady state after "mild" outgasing; this temperature is sufficient to remove a water film but hopefully dos not affect the surface;
3   300K, slow cooling down, cryogenic pump is filled with LN&LN
4   77K ,UCN cryostat filled with LN, cryogenic pump is filled with LHe&LN,
5   4K; UCN cryostat filled with LHe&LN, cryogenic pump is filled with LHe&LN,
6   77K, UCN cryostat filled with LN&LN, cryogenic pump is filled with LHe&LN
7   170K, UCN cryostat 4K bath is empty, 77K bath is filled with LN, cryogenic pump is filled with LHe&LN.

In the second run steps 5 and 6 were repeated 2 times more to check the reliability of the data, while in the first run step 7 was skipped. In the first run we were adjusting our intervals for filling, cleaning, storing and emptying bottle while the cryostat was cooling down to 77K. Therefore only the low temperature data of Run 1 (77K and 4K points, see below) maybe used for physical conclusions.

## Experimental results

The technical test of the UHV cryostat was successful. The vacuum conditions during data taking were <$10^{-8}$ mbar in the first and <$10^{-9}$ mbar in the second run. In the first run the storage bottle was easy cooled down to 4K. The only serious trouble was that the shutter on the UCN bottle was not sealing the storage volume properly. Consequently, our storage time was rather short compared with the expected value. In run 2 the adjustment of the UCN shutter was improved giving better storage time. The radiation shield of the UCN bottle was simplified to improve the pumping speed. The thermometers were remount and in Run 2 it was measured only 30K at the bottom while the cryostat was filled with LHe. The later test reveal that it was a thermo contact not as good as in the first run giving 20K offset to higher value. Therefore, we are sure that the temperature at the bottom was at least 10K while the top was at 4K.

The raw data for the temperature dependence of the loss rate for both, hydrogenated and deuterated surfaces are shown at the Fig.2. The upper curve corresponds to the first run, when storage time was shorter and the surface contains more hydrogen. Two lowest were measured after deuteration of the surface and improvement of UCN shutter. Nevertheless, in the second run during first warming up from 4K up to 77 the shutter again was adjusted giving rise to a slightly better storage time (the lowest short curve). To check the stability we repeated cooling down and warming up between 4K and 77K twice. The results were indistinguishable from each others. It can be considered like an evidence that we do did not have any significant adsorption at the lowest temperatures and our data can be used for a theoretical interpretation. Thus the average values are presented at the lowest curve of Fig.2 to decrease the statistical uncertainties.

Apparently, the statistics could be improved .Nevertheless, we may already conclude that data from both Run 1 and Run 2 clearly demonstrate a well pronounced temperature dependence below 77K with difference only in the amount of Hydrogen.

Another important observation is that our deuteration procedure in $D_2O$ vapour made the Cu surface hydrophobic. Indeed, after deuteration the bottle was exposed to the high humidity air of the experimental hall over night but the loss rates on the non-heated and heated deuterated surface are practically the same. Therefore, the temperature dependence of the second run is free from the contribution of the physisorbed water and may be related to the chemically bound hydrogen.

In several months after the measurement with UCN was done, a chemical content of the copper surface has been studied with the elastic backscattering method (ERDA) at ISL, HMI. It turns out that the Hydrogen content in the bulk (>2$\mu$m) is 0.1% while in the surface layer (7.2·$10^{17}$ at/$cm^2$, ≈ 100Å) it was found to be 8% (5.8·$10^{16}$ at/$cm^2$). The rest was Cu (55%), C(10%), O(25%) with small traces of contamination by Cl,S,Na. The deuterium was found in the amount of 0.6·$10^{15}$ at/$cm^2$ and probably not directly on the surface. The latter conclusion is very approximate because the ERDA technique could not distinguish between deepness and roughness. Our surface was never been polished and visually rough and oxidised after the deuteration. In addition, an accumulation of the contamination from the atmosphere also should affect the results. Therefore, the observed amount of Hydrogen we could consider as the upper limit.

## Data interpretation

For the discussion below we consider only the temperature dependent part of the loss rate $1/\tau(T) = 1/\tau_{exp}(T) - 1/\tau_{exp}(10K)$. To improve statistic both curves of Run 2 were averaged together. The upper limit on amount of the Hydrogen in the top layer of 8% allowed two models to be considered. One is a dilute solution of the Hydrogen in Copper, while another is a hydrogenated film or film clusters on the Cu substrate. In both cases the shape of the temperature dependence arises from the temperature behaviour of the inelastic upscattering cross section while the actual loss rate depends as well on the ratio between the Fermi potential and neutron energy.

If the surface is a compound of several elements, the total the loss rate is a sum of the partial losses. Since for C,O,D and Cu both, coherent and incoherent upscattering cross sections for the temperature range 500K-4K are rather small, the observed value of the losses should be attributed to the incoherent upscattering on the Hydrogen. Indeed, in (n,γ) study [15] a linear correlation between the UCN absorption rate on Hydrogen, $\mu_{cap}$, and the upscattering rate, $\mu_{ie}$ was observed. The $\mu_{cap}$ is proportional to the Hydrogen concentration that was changed by annealing of the sample. Thus the correlation between $\mu_{cap}$ and $\mu_{ie}$ is a direct evidence of the hydrogen origin of the UCN upscattering. Moreover, from the linear fit it was found that

$$\mu_{ie} \approx 17 \mu_{cap} \qquad (1)$$

Since both rates are proportional to the cross sections, the formula (2) allows us to estimate the absolute value of the neutron upscattering on the Cu surface $\sigma_{ie}(H,Cu) \approx 17\sigma_{cap}(H) = 5.6$ b for thermal neutrons. It is an average value for the sample coated with water film and annealed at 750°C. For the samples annealed with an intermediate step the coefficient of the proportionality was changed to 12-15 for all studied samples. It implies $\sigma_{ie} \approx (12-15)\sigma_{cap}(H) = (4-5)$ b per proton. This value might be certainly related to the surface without physisorbed water.

We start from a model of the dilute solution that means a subbarrier UCN interaction with Hydrogen tightly bound to the heavy copper atoms. To calculate the upscattering cross section $\sigma_{ie}(H,Cu)$ we have used a one-phonon approximation, the general formula of which can be written as following:

$$\sigma_{up}(T) = 4\pi b^2 \int e^{-2W(T)} g(\omega,\mu) n(\omega,T) \sqrt{\frac{\omega}{E_0}} d\omega \qquad (2)$$

where $\mu = m/M$, $m$ and $E_0$ is energy and mass of the neutron, $M$ is a mass of scattering nucleus, $n(\omega,T) = 1/(\exp(\omega/T)-1)$, integration is taken over the entire frequency range. $W(T)$ is the Debey-Weller factor:

$$2W(T) = \omega \int_0^{\omega_{max}} \frac{g(\omega',\mu)}{\omega'} \coth\frac{\omega'}{2T} d\omega' \qquad (3)$$

and function $g(\omega,\mu)$ is a generalised density of vibrational states (GDVS) of Hydrogen including the mass factor. Formulas (2) and (3) are derived from the standard expression for the double differential cross section and DW factor (see, for instance, [16]) taking into account the angular isotropy of the upscattered neutrons and the fact that in the case of UCN the initial energy ≈0. Thus both, momentum and energy transfers are equal to the momentum and energy of annihilated phonons, $Q=q_i-q_f \approx q_f$, $\varepsilon=E_i-E_f \approx E_f = \omega = Q^2/2m$. For the discussion below we have emphasized terms dependent on T.

It is known that Hydrogen tightly bound to a lattice undergoes two types of vibrations: acoustical, usually in the energy range up to 40 meV, and optical with higher energies, usually above 100 meV. In the first approximation the shape of an acoustical part mirrors the shape of the phonon spectrum of the lattice. Therefore the phonon spectrum of Hydrogen could be reconstructed a following way

$$g_{H,Cu}(\omega,\mu) = \begin{cases} \delta(\omega - \omega_{opt}), \omega > 40 meV \\ \mu g_{Cu}(\omega), \quad \omega < 40 meV \end{cases} \quad (4)$$

where $\mu=1/65$, $\omega_{opt}$ is a frequency of the optical branch and we assume an equal number of the acoustical and optical branches with normalisation to 1. The phonon spectrum of the copper lattice was calculated in [17] based on experimentally measured force constants. The value of the $\omega_{opt}$ was estimated from CuPdH study to be 116 meV [18]. It is known that the defects could affect the oscillation frequency softening the spectrum. This effect might be well pronounced on the surface. To estimate possible effect we have used 3 different optical frequencies, $\omega_{opt}$=116, 110 and 90 meV. The calculated cross sections are shown on the Fig. 3a.

The absolute value of $\sigma_{ie}(300K)$ calculated with $\omega_{opt}$=116 meV is about 1.2 barn that is too small compare with the value 4-5.6 b. Moreover, the T-dependence raises much steeper than the experimental data. We could see it by the ratio $\sigma_{ie}(300K)/\sigma_{ie}(77K) \approx 13$ in contrary to the experimental $\eta_{exp}(300)/\eta_{exp}(77K) \approx 7$. Taking softer optical frequency of 90 meV, which is probably too extreme, we could raise $\sigma_{ie}(300K)$ up to 2 barn but without any change at temperatures below 100K. Thus, the ratio $\sigma_{ie}(300K)/\sigma_{ie}(77K)$ becomes even larger – up to 40. Softening of the lattice part should also take place but its contribution is much smaller.

Now let us discuss the probability of the UCN subbarrier upscattering per collision, $\eta_{ie}$. The theoretical temperature dependent probability per collision, $\eta(T)$, for sub-barrier UCN losses can be calculated from the $\sigma_{ie}(T)$ using optical potential approximation like:

$$\eta(T) = \frac{\sum_i c_i \sigma^i_{up}(T)}{2\lambda \sum_l c_l b_l} \approx c_H \frac{\sigma^H_{up}(T)}{2\lambda <b>} \quad (5)$$

where $\sigma^j_{up}(T)$ is a partial upscattering cross section for the neutron wavelength $\lambda$, $b_i$ is a scattering wavelength and $c_i$ is a partial concentration of the $i$- nucleus [2,24]. We took into account that the denominator is describing the average scattering length of the surface layer, $<b>$=6.09, calculated for homogeneous mixture of Cu, O, C and H. The experimental values were derived from the raw data a usual way

$$\eta_{H,Cu} = \frac{1}{\tau} \frac{1}{\langle vf \rangle}, \quad (6)$$

where $\tau$ is a storage time, $v$ is a frequency of collision, $f$ takes into account averaging of the isotropic flux over the incident angles. The latter term $<vf>$ was calculated to be approximately $<vf>$=65 s$^{-1}$ using mean UCN velocity averaged over the energy range from 50neV (lower energy cut off by Al foil) up to 165 neV (higher energy cut off by Cu walls) and geometrical area of the bottle $S$, $v \propto S$. In turns, the surface area of a storage bottle depends on the surface roughness. Thus, the absolute values of both, $<\eta_{H,Cu}>$ and $<v>$ are correct only within in a factor. Since our bottle had badly rough, unpolished surface, we include an additional factor 2 to account for the roughness that does not affect the temperature dependence itself.

Experimental probability derived from the data of Run 2 is shown at Fig. 3b together with the Cu-H model calculations. It is clear that the loss probability $\eta_{cal}(T)$ calculated using formula (1) with 8% concentration of Hydrogen is of order of magnitude below experimental values. Thus this model cannot explain neither the shape of the temperature dependence, nor the amplitude of the upscattering probability per collision.

To understand what a frequencies in the spectrum of vibrations could reproduce the low temperature part let us studied a general behaviour of the upscattering cross sections. We choose an exact solution of the single harmonic oscillator that was first used by Weinstock as start point for his pioneer one phonon calculations in 1948 [19]. Another deriving of the solution readers can find in [16,20]. In this model instead of integral we have sums over all possible transitions from state $l=1,..n-1$ to $n=1,2,....N$ with the energy and momentum transfers:

$$\varepsilon = E_i - E_f \approx E_f = (l-n)\hbar\omega_0 = Q^2/2m \qquad (7)$$
$$Q = \sqrt{2m\varepsilon} = \sqrt{2m\hbar\omega_0(l-n)}.$$

The upscattering cross section could be then written as

$$\sigma_{up}(T) = \sigma_{inc}^2 \mu \sqrt{\frac{\omega_0}{E_{th}}}(1-e^{-1/x})\sum_{n=1}^{\infty} e^{-n/x}(n!)\sum_{l=0}^{n-1}(\sqrt{n-l})l!e^{-(n-l)\mu}(\mu(n-l))^{(n-l)}F_{n,l};$$

$$F_{n,l} = \sum_{k=0}^{l}\frac{[-(n-l)\mu]^l}{k!(l-k)(n-l+k)} \qquad (8)$$

where $\mu=m/M$, $x=T/\omega_0$ and we took into account that product of an amplitude of the zero-point oscillation, **b**, and momentum transfer $b^2Q^2 = Q^2/M\omega_0 = 2\mu(n-l)$ ; the incoherent cross section $\sigma_{inc}=4\pi(b_i)^2$.

The results of our calculations together with the experimental points are shown at Fig.4. It is easy to see that to explain the shape of the low temperature part we need rather low oscillation frequencies below 10 meV. Moreover, the peak should be strong enough to give contribution into the cross section, which is comparable with the higher frequencies. Such spectra are rather unusual for metal hydrides but typical for the intermolecular vibrations of molecules weakly bound to a crystal. Evidently, Hydrogen should be bound to rather light atoms to avoid the suppression of the low temperature part by µ factor. It could be either Carbon or/and Oxygen always found in the top layer.

Thus, we naturally arrive to another, more realistic model of the UCN upscattering on the surface that is neutrons upscattered by a hydrogenated film or surface clusters with Fermy-potential close to zero. In that case neutron's interaction obey 1/v low and even small amount of the hydrogen could imply a significant upscattering rate. A similar situation was observed in the stainless steel with UCN capture by Ti nuclei. 1% of Ti gave rise the same capture rate as 50% of Fe due to the presence in the clusters with $V_F$=36 neV in contrary to the average $V_F$=185 neV of the stainless steel[21].

The light hydrogenated molecules always present on the surface are, for instance, water. It turns out that the generalised density of states $g(\omega,\mu)$ of the ice do has a strong low energy band with a peak at 7 meV[22]. To calculate the cross section we still could use formula (3) neglecting a multi-phonon contribution that is about few percent at 300K[23].

Both, the density of state $g_{ice}(\omega,\mu)$ and $\sigma_{ie}$ are shown at Fig.5a. The temperature dependence looks linear-like and $\sigma_{ie}(300K)\approx 4$ b. This value is in a good agreement with $(n,\gamma)$ estimation and the low temperature part is in exellent agreement with the data

Now let us estimate the loss probability using a simplest model of the monoenergetic neutron interacting with a film $V_F < E_{UCN}$ on the substrate with $V_F > E_{UCN}$. As it was shown above the temperature dependent contribution from the H-Cu can be neglected. The absorption in Cu is temperature independent. Therefore, we could neglect interaction with the substrate and write the probability of the upscattering in the film as following

$$\eta_{film} = \sigma_{ie} N [2d + I(\lambda/d, E/V)], \qquad (7)$$

where $N$ is a volume density of Hydrogen nuclei, $d$ is a film thickness, $\sigma_{ie}$ is a cross section of the neutron with energy $(E_{UCN}-V_F)$ in the film, $I(\lambda/d,E/V)$ is an oscillating interference term and. Since the latter must be averaged over the total surface with various $d/\lambda$, we could neglect it and use the formula

$$\eta_{film} = 2\sigma_{ie} n \qquad (8)$$

where we replaced the term $Nd$ by a surface density $n$. Thus, in this model the loss rate is independent on the incident angle and equal to $1/\tau = \eta_{film} <v>$. As a result, $\eta_{film}$ is larger than probability $\eta_{H,Cu}$ derived using the sub-barrier model since here we don't have the factor $<f>=1.6$. Again, we see a quite good agreement between the calculated and experimental curves at Fig.5b.

## Summary

We report about the first study of the temperature dependence of the ultracold neutron's upscattering carried out under ultrahigh vacuum conditions and temperatures range down to 4K. 2 states of the surface of copper bottle were studied – after ultrasonic wash in the distil water and after chemical cleaning and deuteration in the heavy water vapour. Since the latter procedure made surface hydrophobic and we carried out heating of the sample up to 450K we can attribute the T-dependence to the chemically bound Hydrogen.

The main observation of our experiment is a linear-like temperature dependence of the upscattering cross sections in the whole temperature range including the interval between 4K and 77K for both samples. For the deuterated surface the low temperature measurement was repeated twice. We did not see any differences in the data due to cooling or warming. Thus we have demonstrated experimentally that there could be still significant upscattering rate at the liquid nitrogen temperature, $\eta\approx 2.5\cdot 10^{-5}$, in contrary to a long time faith that cooling down to 77K is sufficient for upscattering into meV range to be frozen. It means that an extrapolation of the upscattering rate observed at 77K to 4K point made in [8] could be not valid. It is interesting, that authors derived value $\eta(77K)\approx 2.5\cdot 10^{-5}$ for the rolled Be foil that is very close to ours. The lesson is that only a complete measurement down to at least 4K could put a upper limit on the upscattering rate.

We have analysed our experimental data using an independent measurement of the surface Hydrogen content (8%, $n=5.6\cdot 10^{16}$ cm$^{-2}$), data of prompt $(n,\gamma)$ surface study, $\sigma_{ie}(300K)\approx(4-5)$b, and the phonon data (Cu, Cu-H, ice) from the neutron scattering. Two basic models were considered, hydrogen dilute solution in the metal lattice (H-Cu, subbarrier UCN interaction) and hydrogenated film ( 1/v low).

The information about the Hydrogen surface density allowed us estimate the contribution of both models to $\eta_{exp}(T)$ and to rule out reliably the model of the subbarrier interaction. The absence of such data was a trouble for all previous studies. Only the model of the surface hydrogenated clusters or film with low Fermi-potential could explain the loss probability. In turns, the analysis of the low temperature shape of $\eta(T)$ and comparison of the calculated $\sigma_{ie}$ with (n,γ) data allowed us to draw conclusion that ,again , it can not be film of the H-Cu compound but Hydrogen should be bound to a light mass like Carbon or Oxygen.

The latter conclusion has been already drawn in our first presentation [25] together with an assumption that C-H or O-H based molecules should have a pronounced low frequency's vibrations. Since then we analysed available neutron scattering material about hydrogenated molecules. The intramolecular frequencies usually lie above 100 meV while intermolecular, i.e. translational and librational modes, are below 40 meV. The real issue was to satisfied several conditions simultaneously: a strong peak with energy ≤10 meV, a light weight and formula $X_aH_b$ with b=2a to provide a negative potential and a≤2 for a molecule to be light. For instance, low frequencies could be easily found in the heavy aromatic molecules[26] but then the amplitude of vibration is small due to a mass factor. Another requirement would be a common presence on the surface in the form of clusters.

Finally, we succeed to find a magic molecule. It turns out that usual ice have a remarkable strong translational branch with a low energy peak at 7 meV. The molecule is light, has negative Fermi potential, commonly presents on the surface and tends to adsorb preferably in clusters onto metal surfaces [27]. The calculated value of $\sigma_{ie}(300K)_{ice}$ turns out to be ≈4b that is in a perfect agreement with (n,γ) estimation. Loss probability $\eta_{ice}(T)$ calculated in the film model using the cross sections $\sigma_{ie}(T)_{ice}$ and measured H-density looks surprisingly similar to the experimental $\eta_{exp}(T)$. If ice in the clusters is partly amorphous then the frequencies would be washed out towards lower energies [28]. Thus, presence of ice clusters, which could be reached by UCN without intermediate barriers, could explain quite well both, T-dependence and the loss probability value.

The important consequence is that even a small contamination of the physisorbed water could give rise to the low temperature loss rate. The first storage curve in [4] before annealing could be a case. It is easy to find the water before annealing but more difficult to explain the presence of the water afterwards. A possibility would be that water clusters preferably grow at the bottom of cracks or the valleys between metal grains. A direct correlation between growths of both, clusters of water and $Cu_2O$ have been observed in [29]. A very thin oxide film could seal the clusters being transparent for UCN.

Another possible but more exotic explanation would be $H_2$ babbles like in the bulk of Cu. The first rotational state of $H_2$ molecule has energy ≈15 meV, i.e. 150K [33]. Since the solubility of hydrogen is extremely low, the bulk hydrogen forms babbles. The effect is well known in cryogenic. The density of the bulk hydrogen in our sample was found to be only 0.1% at the depth > 1 μm. One can imagine the increase of the babbles density towards the surface. But then neutron should penetrate through the Cu walls thick enough to hold $H_2$ with a higher potential.

The water-like layer as a possible cause of UCN upscattering has been recently considered by A. Steyerl [30] in connection with the UCN storage on Fomblin oil. He showed that reasonable concentration of H could explain the experimental data only in 1/v model. The contradiction is that Fomblin oil is known to soak in the contaminations, the capability that provides excellent UCN storage properties even at room temperature. Much recent results and their analysis published in [10] allowed authors to distinguish the surface effect from the bulk. Thus, the larger part of the room temperature losses were addressed to the surface. Theoretically the latter was studied in [31] using experimental data about the surface waves. The agreement is quite good. At low temperatures the surface effect became frozen and the bulk losses are in good agreement with subbarrier calculations based on use of the inelastic cross section measured by transmission of 9 m/s neutrons. The effect of the bulk calculated for the pure polymer is much smaller than experimental $\sigma_{ie}$ [32]. Compare with our subbarrier calculation ($\eta(77K)\approx2.5\cdot10^{-5}$, $n_H \leq 8\%$) we can estimate that admixture as small as ≤1% of the weak bound hydrogen in the bulk can contribute up to $\eta\approx0.5\cdot10^{-5}$ at 77K.

Low frequency excitations are known to be present in the amorphous matter but the relative contribution is not large in general[28,34]. To find out the real cause of the low temperature upscattering in different materials and work out a reliable quantitative model of UCN upscattering we have to carry out an experimental study of samples with a known hydrogenated layer and use the (n,γ) technique to measure the cross section per proton. For instance, it could be a polymer film with the spectrum of excitations well studied.

We would like to emphasize the need of the in-situ monitoring of surface content by (n,γ) technique. Otherwise one can guess anything but reliable prove nothing. In our present estimations we have used only the upper limit of the hydrogen content, but we already could step much further in the analysis than all previous experiments. Nevertheless we still don't know the actual density during data taking. In addition, we are constrained by the approximate knowledge of the frequency of collision which depends on the roughness. We can only guess the value of the upscattering cross sections without direct (n,γ) measurement.

In turns, the theoretical technique to calculate the total cross sections also should be tested further for the case of inelastic upscattering of ultra cold neutrons. Our estimations was made using the first Born approximation. It works well for the wide range of the neutron wavelengths down to cold (20A) whereas for the energies comparable with the Fermi-potential it is not valid in general [13]. The corrections due to re-scattering could increase the UCN sensitivity to the low frequency excitations.

The development of the reliable theoretical description is of a great importance to make progress in both, ultracold neutron storage technique at low temperatures and development of applications to the solid state and surface studies using new generation UCN sources. In present work we have demonstrated a high sensitivity of the upscattering cross section to the lowest cut off of the frequencies and sensitivity to an inelastic signal from a hydrogenated film ≈ 10 nm. This opens an interesting opportunity to use UCN for unique studying of low frequency excitations in the nanometers thick film with $V_F<E_{UCN}$ (polymers, deposited ice and another gases) using both, prompt gamma analysis to monitor Hydrogen amount on the surface and measurement of the upscattering rate at different temperatures by the (n,γ) and the storage technique.

The study of $\sigma_{ie}(T)$ with UCN is similar to the study of the temperature dependence in the specific heat measurement but with a contrast to Hydrogen excitations and in the very thin surface layer. The anomalies found in the bulk $C_V$ (for instance, "boson" peak around 2-4 meV and mK anomaly in polymers) are still under intensive study by various modern methods.

The lowest cut-off of the sensitivity of the existing inelastic scattering instruments is restricted by the elastic peak. In our case an elastic reflection is invisible and we can detect relative signal inel/elast≈$10^{-6}$. At reflectometer the background from the elastic reflection of the substrate would be a serious issue. Inelastic reflectometry simply does not exist at present while UCN seems to be a native tool for such study. Another restriction lies in the thickness of the sample. Typical NIS sample thickness is of order of 0.1-1mm that means bulk samples. Nanoscale film could be studied only with electrons, which have sensitivity to another parameters compare with neutrons. The sensitivity to Hydrogen with the new high density UCN sources (for instance, [35]) could be as good as $10^{15}$ atoms/cm$^2$ .The limitation would rather come from an ambient background.

## Acknowledgements

We would like to thank the TU Munich for support and the design and construction of the cryostat; the technical services of HMI and ILL, especially B.Uhrban(HMI) and T.Brenner and the reactor division (ILL) for the excellent technical support during the experiment. The discussions with A. Steyerl and V.Morosov were especially fruitful and stimulating. Many thanks for A. Kolesnikov for providing his phonon data of ice.

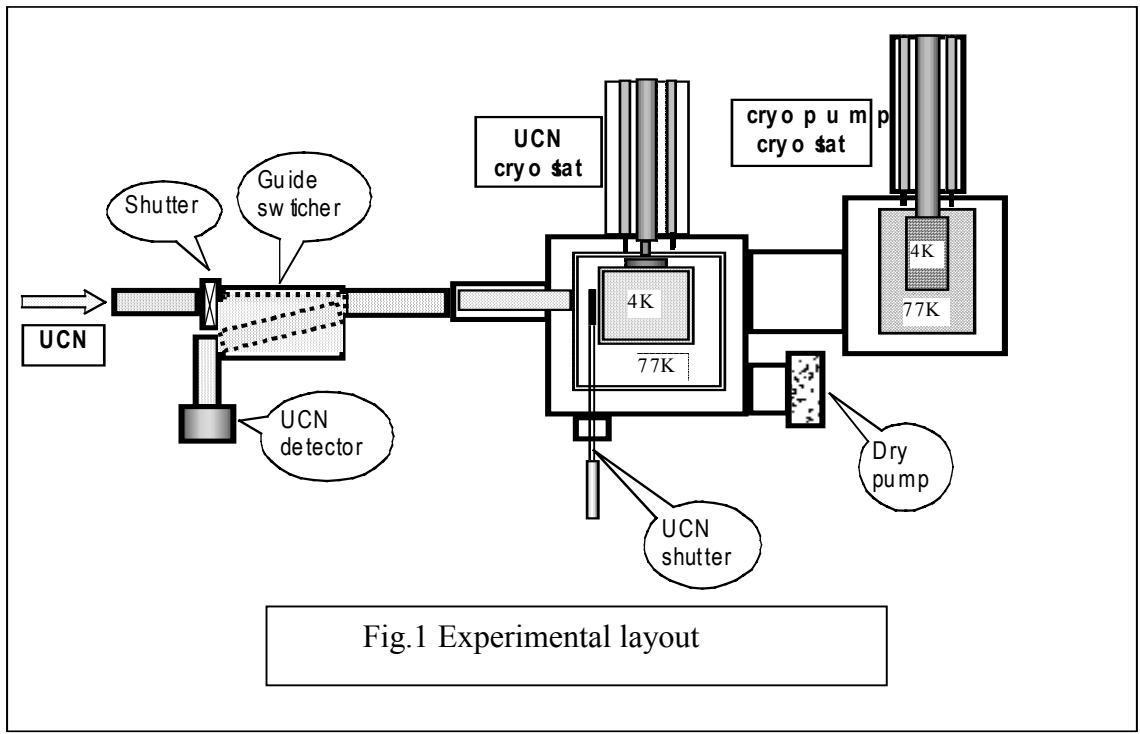

Fig.1 Experimental layout

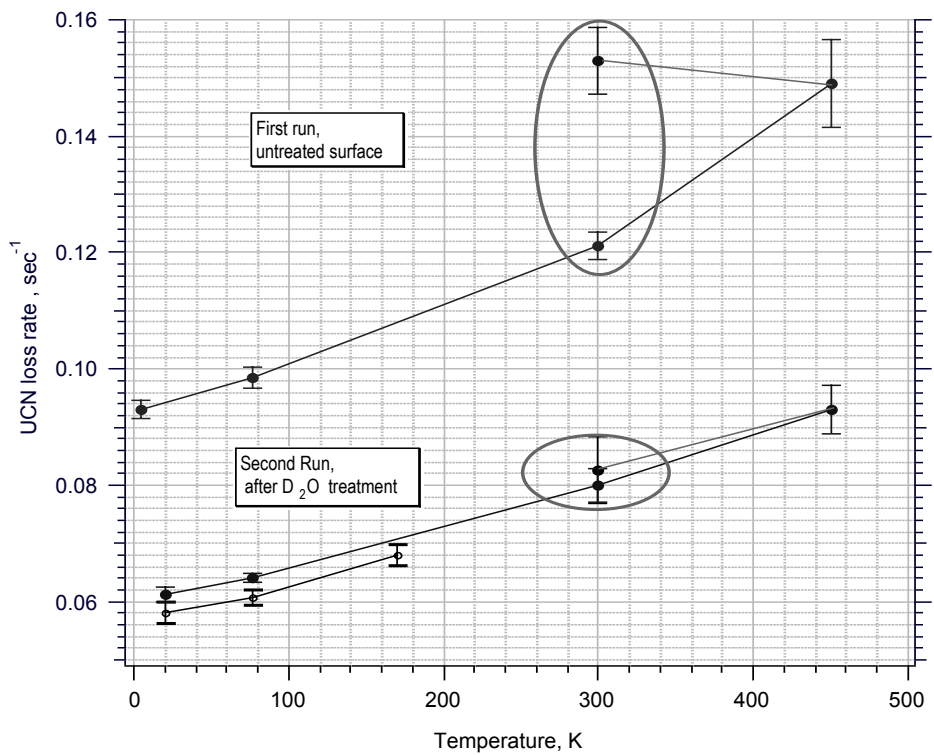

**Fig 2** Raw data for both runs: the upper curve-first run the surface was washed in the ultrasonic bath with the distilled water; the lower two – second the surface was heated in the D$_2$O vapor. The ovals show the comparison for 300 K data before and after heating for two states of the surface.

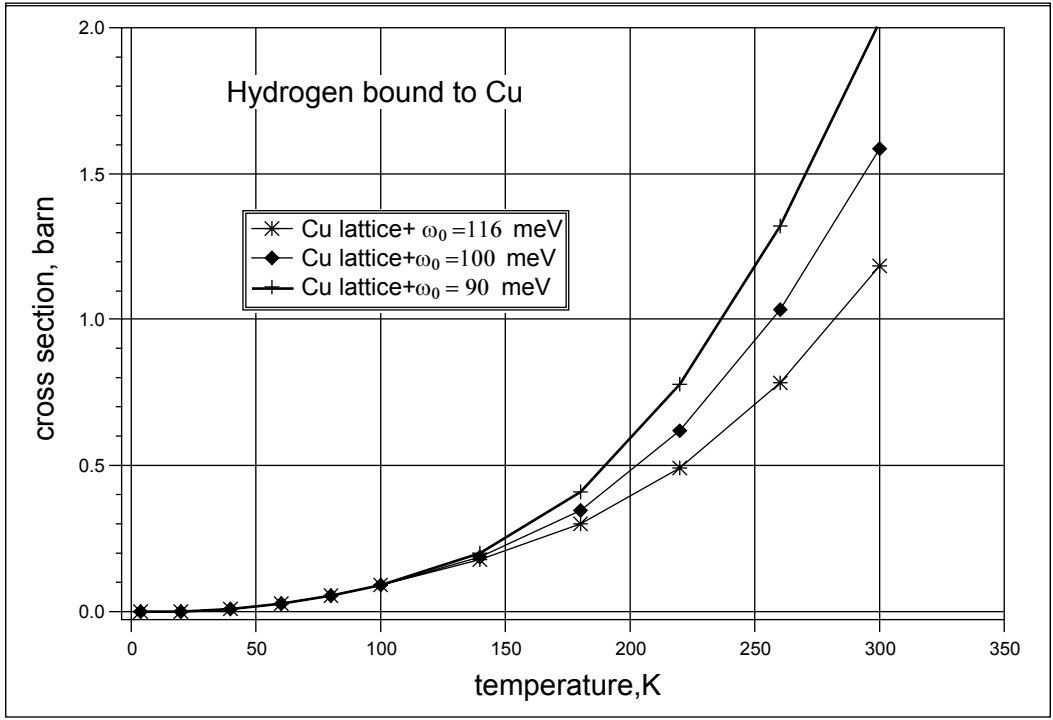

Figure 3a. The upscattering cross sections calculated in the one phonon approximation with 3 different $\omega_{opt}$=116,100 and 90 meV. The lattice part contribute below 100K while optical mode affects strongly 300K value.

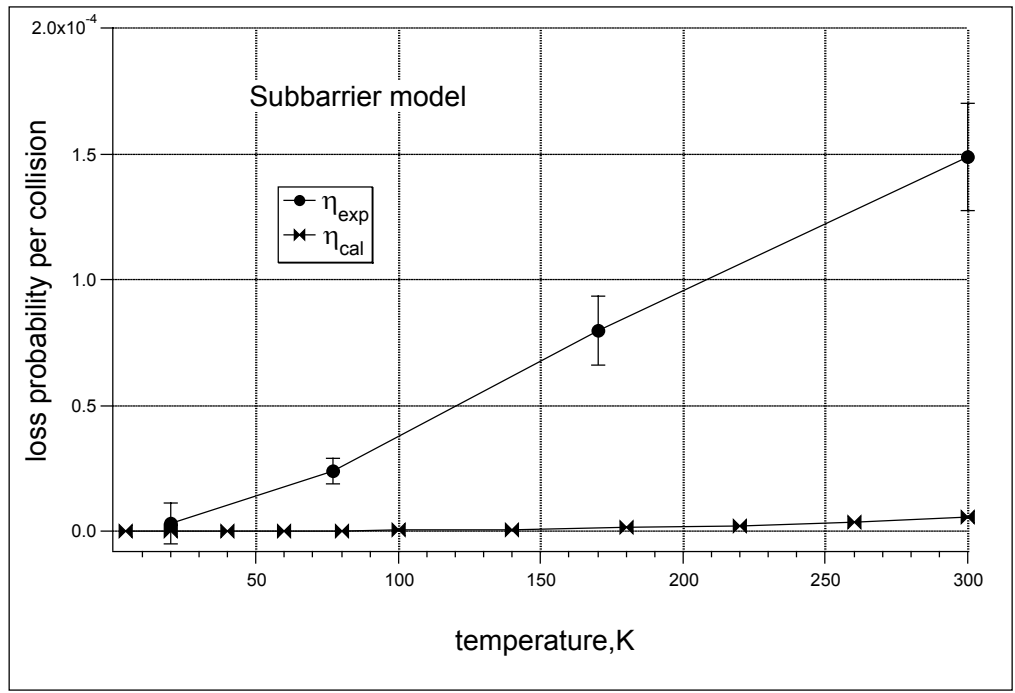

Fig.3b Upscattering probabilities per collision for subbarrier model: $\eta_{exp}(T)$ is derived from experiment and $\eta_{cal}(T)$ is calculated using $\sigma_{ie}(T)$ for Hydrogen bound to Cu.

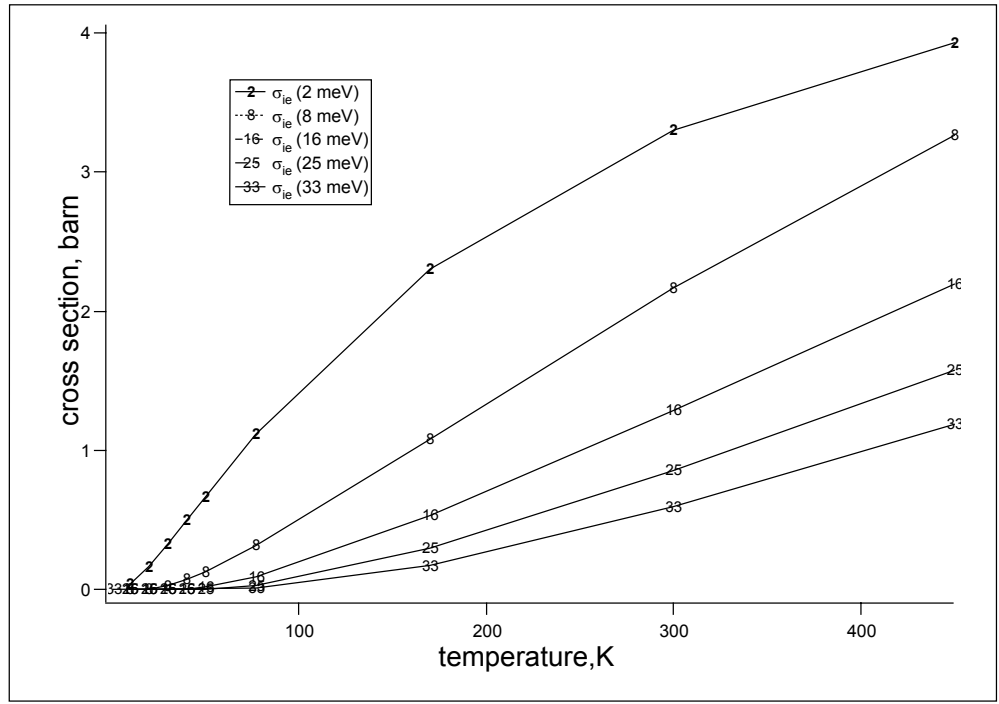

Figure 4. Temperature dependence of the upscattering cross section calculated for harmonic oscillators with frequencies 2, 8, 16, 25 and 33meV and mass factor μ=1/7

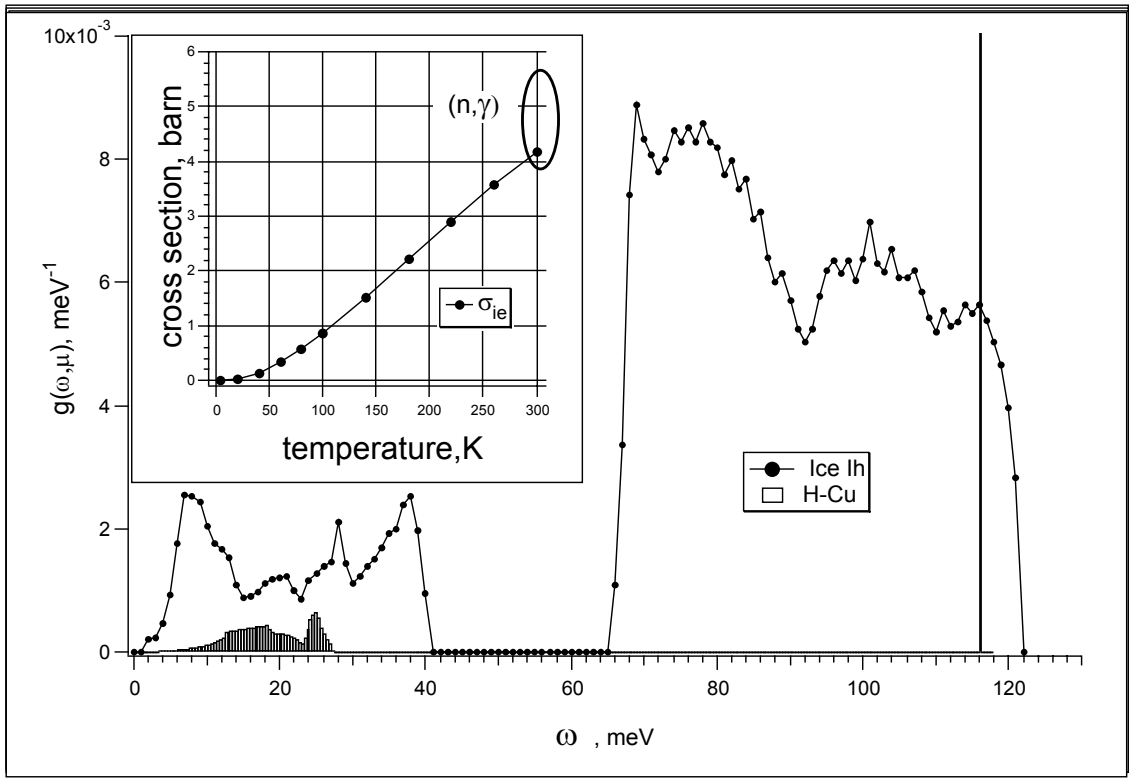

Fig. 5a. Generalized density of states $g(\omega,\mu)$ for Cu(calculated) and Ice Ih (experimental data,[22]). Insert shows $\sigma_{ie}(T)$ calculated using $g(\omega,\mu)$ of ice. The oval shows the range of $\sigma_{ie}(300)$ estimated from $(n,\gamma)$ study.

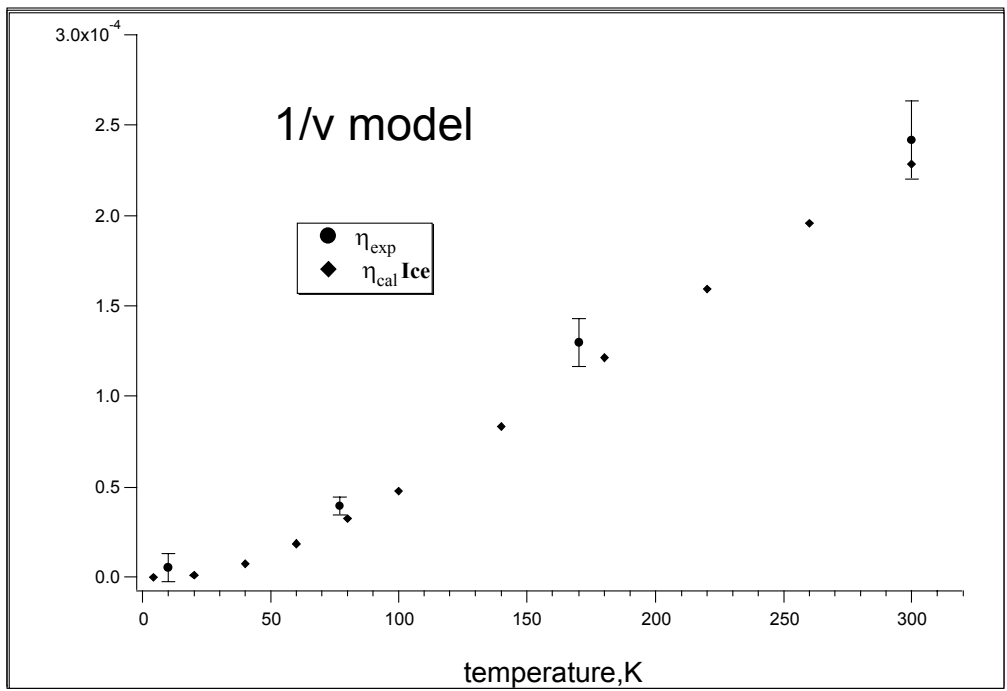

Fig. 5b. Loss probability per collision calculated with $\sigma_{ie}(T)$ of ice and derived from experiment for the film model.